\documentclass{elsart1p}
\usepackage{graphicx}
\usepackage{amssymb}
\def\ignore#1{{}}
\newcommand{\beeq}{\begin{equation}}
\newcommand{\eneq}{\end{equation}}
\newcommand{\beqn}{\begin{eqnarray}}
\newcommand{\eeqn}{\end{eqnarray}}

\def\la{\raise.16ex\hbox{$\langle$}\lower.16ex\hbox{}  }
\def\ra{\, \raise.16ex\hbox{$\rangle$}\lower.16ex\hbox{} }
\def\go{\rightarrow}

\def\onehalf{ \hbox{${1\over 2}$} }

\def\eff{{\rm eff}}

\def\KK{{\rm KK}}

\def\psibar{ \psi \kern-.65em\raise.6em\hbox{$-$} }
\begin{document}
\begin{frontmatter}
%
%
%
\title{Dynamical Electroweak Symmetry Breaking in $SO(5) \times U(1)$ Gauge-Higgs
Unification in the Randall-Sundrum Warped Space}
%
%
\author{Yutaka Hosotani}
\address{Department of Physics, Osaka University, Toyonaka, Osaka 560-0043, Japan}
\begin{abstract}
In the gauge-Higgs unification scenario the Higgs field is unified with gauge fields
in higher dimensional gauge theory.  The 4D Higgs field $H(x)$ corresponds to 4D
fluctuations of the  Aharonov-Bohm phase (Wilson line phase) 
$\theta_H$ in the extra-dimension.
An $SO(5)\times U(1)$ gauge-Higgs unification model in the Randall-Sundrum 
warped spacetime with top and bottom quarks is presented.   
Gauge couplings of the top quark multiplet induce electroweak symmetry breaking
by the Hosotani mechanism.  The effective potential $V_\eff (\theta_H)$ is
found to be minimized at $\theta_H = \onehalf \pi$ and
the Higgs mass is predicted around 50 GeV.  
The $ZZH$ and $WWH$ couplings   vanish at $\theta_H = \onehalf \pi$
so that the LEP2 bound for the Higgs mass is evaded.
The result is summarized in the effective interactions for 
$\hat \theta_H(x) = \theta_H + H(x)/f_H$. 
\end{abstract}
\begin{keyword}
gauge-Higgs unification \sep electroweak symmetry breaking \sep Hosotani mechanism 
%
\PACS 11.10.Kk \sep  11.15.Ex \sep 12.60.-i \sep 12.60.Cn
\hfill OU-HET 621/2009
\end{keyword}
\end{frontmatter}
%

\section{Introduction}
\label{}

In the standard model  the electroweak (EW) 
symmetry is spontaneously broken by the Higgs mechanism, 
which is yet to be  confirmed by experiments.  
Although the Higgs particle  is expected to be found at LHC in the near future,  
it is not clear whether or not the Higgs particle
appears as described in the standard model.  

Recently there has been significant progress in the gauge-Higgs unification scenario
in which the 4D Higgs field is identified with a part of the extra-dimensional
component of gauge fields in higher dimensions. 
The Higgs field appears as an Aharonov-Bohm (AB) phase, or a Wilson line phase,
in the extra dimension,  thereby  the EW symmetry being dynamically broken 
by the  Hosotani mechanism.\cite{YH1,  Davies1}
One of the promising models is
the $SO(5)\times U(1)_X$ gauge-Higgs unification model in the Randall-Sundrum
(RS) warped spacetime, in which many  
predictions are given.\cite{Agashe1}-\cite{HK}

\section{Model}

The metric of the RS warped spacetime is given by 
$ds^2 =z^{-2} ( \eta_{\mu\nu} dx^\mu dx^\nu +k^{-2} dz^2)$,
$1\le z \le z_L = e^{kL}$. The  bulk region $1<z < z_L$ is the AdS space with 
a cosmological constant $\Lambda = - 6k^2$,
which is sandwiched by the Planck brane at $z=1$ and the TeV brane at $z= z_L$.
We consider an $SO(5)\times U(1)_X$ gauge theory  in this space.\cite{HOOS}
The orbifold boundary conditions
at the two branes break the symmetry to $SO(4)\times U(1)_X$.  Additional brane
interactions on the Planck brane further break it down  to  the EW symmetry 
$SU(2)_L\times U(1)_Y$.  
The 4D $SU(2)_L$ doublet Higgs field resides in the $(a5)$-components 
($a=1,\cdots, 4$) of  $A_z(x,z)$.  
The Aharonov-Bohm (AB) phase  is given by the phase of 
$(P \exp \big\{ i g \int_{1}^{z_L} dz  \,  A_z \big\})^2$.
$A_z^{(45)}$ contains the 4D neutral Higgs field $H(x)$.  

Fermions are introduced in the vector ({\bf 5}) representation of $SO(5)$
in the bulk. In terms of $SU(2)_L$ content they are, for the top and bottom multiplets, 
\beeq
\Bigg( \bigg[ \begin{array}{c} T \\ B \end{array}\bigg] ,
         \bigg[ \begin{array}{c} t \\ b\end{array}\bigg]  ,  t'  \Bigg)
~~,~~
\Bigg( \bigg[ \begin{array}{c} U \\ D \end{array}\bigg] ,
 \bigg[ \begin{array}{c} X \\ Y\end{array}\bigg]  ,  b'  \Bigg) ~~.
 \label{bulkF}
\eneq
In addition, right-handed brane fermions in the $({\bf 2}, {\bf 1})$ representation of 
$SU(2)_L \times SU(2)_R$ 
are introduced on the Planck brane.  
At low energies the desired mass spectrum for the top and bottom quarks is
obtained.  All other fermions become heavy.  

When the AB phase $\theta_H$ associated with $A_z^{(45)}$ becomes
non-zero, the EW symmetry is broken, thereby $W$, $Z$, top and bottom quarks 
acquiring masses.  In particular, 
\beeq
m_t \sim
\frac{m_\KK}{\sqrt{2}\pi} \, \sqrt{1 - 4c^2} ~  |\sin\theta_H| ~~,~~
m_b \sim \Big| \frac{\tilde \mu}{\mu_2} \Big| ~ m_t
\label{topbottom}
\eneq
where the Kaluza-Klein mass scale is given by $m_\KK \sim \pi k z_L^{-1} \sim 1.5 \,$TeV.
$c$ is the bulk mass parameter ($\sim 0.43$).  $\tilde \mu$ and $\mu_2$ are mass 
couplings between the bulk  and brane fermions.

\begin{figure}[t,b]
\centering  \leavevmode
\includegraphics[height=3.8cm]{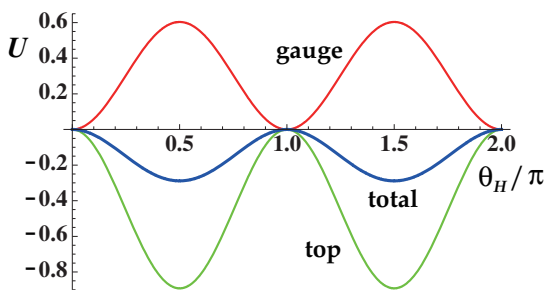}
\caption{The effective potential $V_\eff (\theta_H)$.
$U(\theta_H/\pi) = (4\pi)^2 (kz_L^{-1})^{-4} V_\eff (\theta_H) $ is plotted
for $z_L = 10^{15}$.}
\label{effV-fig}
\end{figure}

\section{Dynamical EW symmetry breaking}

The effective potential $V_\eff (\theta_H)$ is evaluated at the one loop 
level.\cite{MSW, HOOS, Oda1, Falkowski1, Hatanaka1}
The result is depicted in fig.\ 1.  Without the fermions  $V_\eff (\theta_H)$
is minimized at $\theta_H = 0$ or $\pi$ so that the EW gauge symmetry is unbroken.
With the contributions from the top quark multiplet, $V_\eff$
has global minima at $\theta_H = \pm \onehalf \pi$ so that the EW gauge symmetry 
is dynamically broken.  Contributions from light quarks ($c,s,u,d$) are negligible.
In the flat spacetime limit ($k \go 0$) the symmetry remains unbroken.

The Higgs mass is given by
\beeq
m_H^2 =  \frac{1}{f_H^2}  \, 
\frac{d^2 V_\eff}{d \theta_H^2} \bigg|_{\rm min} 
~~,~~
f_H \sim \frac{2}{\sqrt{kL}} \frac{m_\KK}{\pi g}~.
\label{Higgs3}
\eneq
$m_H$ is predicted around 50 GeV for $z_L = 10^{13} \sim 10^{17}$.  
We stress that this is in no conflict with the  LEP2 bound for $m_H$.  
The $ZZH$ coupling vanishes at  $\theta_H = \onehalf\pi$.
The process $e^+ e^- \go Z \go Z H$ does not take place 
so that the LEP2 bound is evaded.

\section{Effective interactions}

The Higgs interactions are summarized in the effective interactions\cite{Sakamura1, HK}
\beqn
&&
{\cal L}_\eff  = - V_\eff (\hat \theta_H) - m_W(\hat \theta_H)^2 W^\dagger_\mu W^\mu
- \onehalf m_Z(\hat \theta_H)^2 Z_\mu Z^\mu  
- \sum_f m_f(\hat \theta_H) \psibar_f \psi_f ~, \cr
\noalign{\kern -3pt}
&&
m_W(\hat \theta_H) \sim \cos \theta_W \, m_Z(\hat \theta_H)
\sim \onehalf g f_H \sin \hat \theta_H  ~, \cr
\noalign{\kern 3pt}
&&
 m_f(\hat \theta_H) \sim \lambda_f \sin \hat \theta_H
 ~~,~~
\hat \theta_H (x) = \theta_H + \frac{H(x)}{f_H} ~~.
\label{effective2}
\eeqn
The first term is the effective potential for $\hat \theta_H$.  
$m_W(\hat \theta_H)$ and $m_Z(\hat \theta_H)$ are given in \cite{SH1}.
$m_W = m_W(\theta_H)$, $m_Z = m_Z(\theta_H)$.   
Expanding $m_W(\hat \theta_H)^2$ and 
$m_Z(\hat \theta_H)^2$   in a power series in $H$, one finds
that the $WWH$ and $ZZH$ couplings are suppressed by a factor
$\cos \theta_H$ compared with those in the standard model.
For the $WWHH$ and $ZZHH$ couplings the suppression factor becomes
$\cos 2 \theta_H$.  
$m_f(\hat \theta_H)$ is given in \cite{HK}.
It follows that the $WWH$, $ZZH$ and Yukawa couplings vanish at 
$\theta_H=\onehalf \pi$, which leads to new phenomenology.

\vskip -.5cm 


%
%
\def\jnl#1#2#3#4{{#1}{\bf #2} (#4) #3}

\def\Zphys{{\em Z.\ Phys.} }
\def\jssc{{\em J.\ Solid State Chem.\ }}
\def\jpsJ{{\em J.\ Phys.\ Soc.\ Japan }}
\def\ptps{{\em Prog.\ Theoret.\ Phys.\ Suppl.\ }}
\def\PTP{{\em Prog.\ Theoret.\ Phys.\  }}

\def\JMP{{\em J. Math.\ Phys.} }
\def\NPB{{\em Nucl.\ Phys.} B}
\def\NP{{\em Nucl.\ Phys.} }
\def\PLB{{\em Phys.\ Lett.} B}
\def\PL{{\em Phys.\ Lett.} }
\def\PRL{\em Phys.\ Rev.\ Lett. }
\def\PRB{{\em Phys.\ Rev.} B}
\def\PRD{{\em Phys.\ Rev.} D}
\def\PRe{{\em Phys.\ Rep.} }
\def\AP{{\em Ann.\ Phys.\ (N.Y.)} }
\def\RMP{{\em Rev.\ Mod.\ Phys.} }
\def\ZPC{{\em Z.\ Phys.} C}
\def\SCI{\em Science}
\def\CMP{\em Comm.\ Math.\ Phys. }
\def\MPLA{{\em Mod.\ Phys.\ Lett.} A}
\def\IJMPA{{\em Int.\ J.\ Mod.\ Phys.} A}
\def\IJMPB{{\em Int.\ J.\ Mod.\ Phys.} B}
\def\EPJC{{\em Eur.\ Phys.\ J.} C}
\def\PR{{\em Phys.\ Rev.} }
\def\JHEP{{\em JHEP} }
\def\cmp{{\em Com.\ Math.\ Phys.}}
\def\JPA{{\em J.\  Phys.} A}
\def\JPG{{\em J.\  Phys.} G}
\def\NJP{{\em New.\ J.\  Phys.} }
\def\CQG{\em Class.\ Quant.\ Grav. }
\def\ATMP{{\em Adv.\ Theoret.\ Math.\ Phys.} }
\def\ibid{{\em ibid.} }

\end{document}